 \documentclass[preprint]{aastex}

\newbox\grsign \setbox\grsign=\hbox{$>$} \newdimen\grdimen
\grdimen=\ht\grsign
\newbox\simlessbox \newbox\simgreatbox \newbox\simpropbox
\setbox\simgreatbox=\hbox{\raise.5ex\hbox{$>$}\llap
     {\lower.5ex\hbox{$\sim$}}}\ht1=\grdimen\dp1=0pt
\setbox\simlessbox=\hbox{\raise.5ex\hbox{$<$}\llap
     {\lower.5ex\hbox{$\sim$}}}\ht2=\grdimen\dp2=0pt
\setbox\simpropbox=\hbox{\raise.5ex\hbox{$\propto$}\llap
     {\lower.5ex\hbox{$\sim$}}}\ht2=\grdimen\dp2=0pt
\def\simgreat{\mathrel{\copy\simgreatbox}}
\def\simless{\mathrel{\copy\simlessbox}}

 \slugcomment{To appear in The Astrophysical Journal}

\shorttitle{\emph{Chandra} Observation of Circinus}
\shortauthors{Smith and Wilson}

\begin{document}

\title{A Chandra Observation of the Circinus Galaxy}


\author{David A. Smith and Andrew S. Wilson\altaffilmark{1}}
 
\affil{Department of Astronomy, University of Maryland, College Park, MD
20742; dasmith@astro.umd.edu, wilson@astro.umd.edu}


\altaffiltext{1}{Adjunct Astronomer, Space Telescope Science
Institute, 3700 San Martin Drive, Baltimore, MD 21218;
awilson@stsci.edu}

\begin{abstract}

We report on a recent \emph{Chandra} ACIS-S observation of the
Circinus galaxy.  These observations confirm that the nuclear spectrum
results from reflection of a hard X-ray continuum by ``neutral''
matter.  The nuclear X-ray emission is extended by $\sim 60$ pc in the
general direction of the optical ``ionization cone''.  An image in the
Fe K$\alpha$ line has been made and shows that this emission extends
up to $200$ pc from the nucleus.  There is also large-scale X-ray
emission both along and perpendicular to the galaxy disk.  Thermal
plasma models for this extended gas indicate temperatures $kT \sim
0.6$ keV, though cooler photoionized gas is also possible.  The X-ray
emission from gas in the disk is probably associated with the
starburst ring of radius $150$--$250$ pc.  The gas extending $\sim
600$ pc perpendicular to the disk is closely correlated with the
high-excitation optical-line emission.  In addition to its soft X-ray
emission, we tentatively detect a hard component from the gas above
the plane; this hard emission may represent nuclear X-rays scattered
into our line of sight by electrons in the outflowing wind.  Ten
compact sources are found in the central kpc of the galaxy.  The most
luminous has an X-ray luminosity of $\simeq 10^{40}$ erg s$^{-1}$ and
seems to be an X-ray binary in the Circinus galaxy with a black-hole
mass exceeding $80 \, M_{\odot}$.

\end{abstract}


\section{Introduction} \label{sec-intro}

The Circinus galaxy is a large, nearby ($4 \pm 1$ Mpc; Freeman et
al. 1977) galaxy that harbors both a circumnuclear starburst and a
Seyfert 2 nucleus.  Evidence for an obscured Seyfert 1 nucleus is
provided by the finding of a broad ($\rm FWHM \simeq 3300$ km
s$^{-1}$) H$\alpha$ line component in polarized light \citep{oli98}.
The picture of an obscured Seyfert 1 nucleus is also supported by both
the discovery of highly ionized gas extending along the minor axis of
the galaxy, with a morphology that is reminiscent of the ionization
cones seen in other Seyfert galaxies \citep{mar94}, and the direct
X-ray detection of the nucleus through a column density of $\sim 4
\times 10^{24}$ cm$^{-2}$ \citep{mat99}.  The \emph{ASCA} and
\emph{BeppoSAX} spectra below 10 keV shows both a flat continuum and
intense Fe K$\alpha$ line emission, characteristic of the Compton
reflection expected from cold gas illuminated by a hard ($\Gamma
\simeq 1.7$) power-law continuum \citep{mat96}.  Superposed on this
spectrum are a number of emission lines from ``neutral'' and ionized
states of Ne, Mg, Si, S, and Fe K$\beta$ \citep{mat96,gua99}.  The
energies and equivalent widths of the soft (0.5--3 keV) X-ray lines
indicate that they cannot be produced in the material responsible for
the Compton reflection, but must instead originate from highly ionized
gas, possibly photoionized by the Seyfert nucleus \citep{sak00}.  A
recent \emph{Chandra} High Energy Transmission Grating Spectrometer
(HETGS) observation supports a two component, centrally-illuminated
model \citep{sam01b}.  Low ionization lines and the hard reflected
continuum above $\sim 2.5$ keV originate from a cold gas in a compact
region near ($< 15$ pc) the nucleus.  High ionization lines and the
softer, lower energy continuum arise in a highly ionized, more
extended (50 pc) gas (cf. Netzer, Turner \& George 1998).  There is
also large-scale X-ray emission towards the NW, which is roughly
perpendicular to the major axis of the galaxy disk and in the
direction of the high excitation gas evident in the optical images
\citep{sam01a}.  A number of compact X-ray sources are also seen.

Here we report on the results from a long ($\simeq 30$ ksec)
observation of Circinus with the spectroscopic array of the Advanced
CCD Imaging Spectrometer (ACIS-S; Garmire et al. 2000).  These
observations allow a study of the extended X-ray emission and compact
X-ray sources with a higher signal to noise than available to
\citet{sam01a}.  In this paper, we adopt a distance of 4~Mpc to the
Circinus galaxy \citep{fre77}, so $1^{\prime\prime} = 19 \rm pc$.

\section{Observations and Data Analysis} \label{sec-obs}

Since the nucleus of Circinus is known to be a strong X-ray source
(e.g. Matt et al. 1996), we were concerned that ``pile-up'' could
affect the \emph{Chandra} observations.  In order to measure the count
rate of the nucleus, we first obtained a short observation (obsid 355)
of Circinus on January 16, 2000.  Single exposures with a 0.4s frame
time were alternated with three exposures with a 3.2s frame time, the
total integration time being $\simeq 1$~ks. This observation showed
that the peak count rate per pixel for good grades taken in the 3.2s
frame time mode was lower than that taken in the 0.4s frame
time. Further, some 30\% of the level 1 grades at the brightest pixel
in the 3.2s frame time data were bad. Both of these effects indicate
that pile-up is significant at the nucleus in a 3.2s frame time.

In order to obtain a spectrum of the nucleus which is free of pile-up,
as well as a wide field deep observation, the science observations
were taken in two exposures, both on March 14, 2000. Both exposures
were taken with the nucleus of the galaxy at the aimpoint of the ACIS
S3 chip. A short ($\simeq 5$~ks) exposure (obsid 365) utilizing a
subarray window of 128 rows on S3 and nominal frame time of 0.4s, was
taken to obtain an ``unpiled'' spectrum of the nucleus. A longer
($\simeq 25$~ks) exposure (obsid 356) was taken with the default 3.2s
frame time and chips I2, I3, S1, S2, S3, and S4 on, providing a deep
integration of a wide field. We concern ourselves with only the
central region of the galaxy, imaged on chip S3, in this paper.

Data reduction was performed using the latest version (1.1.5) of the
\emph{Chandra Interactive Analysis of Observations (CIAO)} software
together with the level 2 events file, which contains only \emph{ASCA}
grades 0, 2, 3, 4, and 6.  The level 2 events were gain corrected
using calibration files appropriate for our observations.  Times of
bad aspect and high background were removed.  Bad pixels and columns
were excluded.  The total good time intervals were 4913s and 24445s
for the 0.4s and 3.2s frame-time exposures, respectively.

\section{Results} \label{sec-results}

\subsection{The Central Region of Circinus} \label{sec-overview}

A \emph{Chandra} ACIS image was extracted in the 0.5--8 keV band for
the central $1\farcm2 \times 1\farcm2$ ($1.35 \times 1.35$~kpc) region
of the Circinus galaxy (Figure~\ref{fig1}).  The bright source at the
center (source D) is the Seyfert nucleus.  The
\emph{Chandra}--measured position of the nuclear source is
$\alpha_{\mathrm x}(2000) = 14^{\mathrm h} 13^{\mathrm m}
09.\!^{\mathrm s}94$, $\delta_{\rm x}(2000) = -65^{\circ} 20^{\prime}
21.\!^{\prime\prime}0$.  This position is $1.\!^{\prime\prime}6$ away
from the optical position ($\alpha_{\mathrm o}(2000) = 14^{\mathrm h}
13^{\mathrm m} 09.\!^{\mathrm s}7 \pm 0.\!^{\mathrm s}1$,
$\delta_{\mathrm o}(2000) = -65^{\circ} 20^{\prime}
21.\!^{\prime\prime}4 \pm 0.\!^{\prime\prime}7$; Freeman et al. 1977)
and $0.\!^{\prime\prime}2$ away from the radio position
($\alpha_{\mathrm r}(2000) = 14^{\mathrm h} 13^{\mathrm m}
09.\!^{\mathrm s}95 \pm 0.\!^{\mathrm s}02$, $\delta_{\mathrm r}(2000)
= -65^{\circ} 20^{\prime} 21.\!^{\prime\prime}2 \pm
0.\!^{\prime\prime}1$; Greenhill 2001).  The X-ray and radio positions
are thus in agreement.  The inner $31\farcs5 \times 31\farcs5$ ($600
\times 600$ pc) region of the field is shown in more detail in
Figure~\ref{fig2}.  The emission from the nucleus is clearly extended
along the NW--SE direction.  There are a total of 11 sources detected
above a nominal $5\sigma$ threshold within the central $\simeq
1^{\prime} \times 1^{\prime}$ region using the {\sc wavdetect\/}
algorithm (distributed as part of the \emph{CIAO} data analysis
package).  Table~\ref{tbl-1} gives the positions and background
subtracted count rates for each source.  The background was taken from
source free regions surrounding each source.  The count rates are for
the combined 0.4s and 3.2s frame time data, except for sources D, F,
and J, which are affected by pileup in the longer frame time data.  We
discuss the nuclear emission in Section~\ref{sec-nucleus}, the large
scale extended emission in Section~\ref{sec-large_scale}, and the
compact non-nuclear X-ray sources in Section~\ref{sec-compact}.

\subsection{The Nucleus} \label{sec-nucleus}

\subsubsection{Extended Emission} \label{sec-nuc_extended}

A radial profile of the X-ray surface brightness was determined by
extracting, from the unpiled short frame time observation, counts in
circular annuli with widths $1~\rm pixel$ ($0.\!^{\prime\prime}5$)
centered on the nucleus (Figure~\ref{fig3}).  The X-ray emission from
compact sources close to the nucleus was excluded, and the background
was estimated from a circular region $40^{\prime\prime}$ SW of the
nucleus, which is free of large scale extended X-ray emission.  The
telescope PSF at 2~keV was calculated using the program {\sc mkpsf}
and file aciss1998-11-052dpsf3N0002.fits.  We find the number of
0.5--8~keV counts within a radius of $3^{\prime\prime}$ is about twice
that of the normalized telescope PSF, within the same radius.

The broad-band X-ray emission from the nucleus shows a bright core
plus an extension of $\simeq 3$--$4^{\prime\prime}$ in position angle
(p.a.) $\simeq 290^{\circ}$ (Figures~\ref{fig1}, \ref{fig2}).  This
direction of X-ray extent is almost perpendicular to that of the
galaxy disk ($\rm p.a. = 25^{\circ}$; Maiolino et al. 1998).
\emph{HST} optical emission-line images \citep{wil00} show a compact
$\mathbf V$-shaped structure (possibly an ``ionization cone'')
extending $\simeq 2^{\prime\prime}$ to the NW (in $\rm p.a. =
316^{\circ}$) of the nucleus.  The direction of elongation of the
nuclear X-ray emission is close to that of the bright southern edge of
this ``$\mathbf V$'', which lies in $\rm p.a. \simeq 280^{\circ}$.
Thus both the scale and direction of the nuclear X-ray extension are
close to those of the optical emission-line ``$\mathbf V$''.
Observations of Circinus in water maser emission reveal gas both in a
warped, edge-on accretion disk and in outflow from the nucleus
\citep{gre01a}.  The mean p.a. of the maser outflow is $308^{\circ}$,
which is also quite close to the direction of the X-ray extension in
the nucleus.

The distribution of Fe K$\alpha$ line emitting gas was determined by
extracting an image in the 6.2--6.5 keV band, which includes all of
the Fe K$\alpha$ line emission, and subtracting the continuum at each
location.  This continuum was taken to be the average of the counts in
the 5.9--6.2 and 6.5--6.8 keV bands.  This continuum--subtracted line
image was then convolved with a 2-D Gaussian profile of width $\sigma
= 1~\rm pixel$ ($0.\!^{\prime\prime}5$) and the result is shown in
Figure~\ref{fig4}.  The Fe K$\alpha$ emission of the nucleus is
extended by a few arcsecs along the EW (in $\rm p.a. \simeq
270^{\circ}$) direction, and there is a plume of emission extending
some $10^{\prime\prime}$ from the nucleus towards the NW diffuse X-ray
lobe (in $\rm p.a. \simeq 330^{\circ}$; Figure~\ref{fig1}).  The EW
extension lies along the southern edge of the ``ionization cone'' and
the plume follows the eastern edge of the ionized gas $\sim
10^{\prime\prime}$ from the nucleus (see Figure 2 of Wilson et
al. 2000).  This raises the possibility that the extended Fe K$\alpha$
emission arises in neutral gas just outside the ``cone''.  We note
that Fe K$\alpha$ emission is evident in the spectra of the nuclear
extension, and the NW plume, and so these features in the Fe K$\alpha$
image are not artefacts of the continuum subtraction process.

\subsubsection{Nuclear Spectrum} \label{sec-nuc_spectrum}

A spectrum was extracted from an elliptical region, with major axis in
$\rm p.a. = 112.\!^{\circ}2$, and centered on the peak flux of the
nucleus.  The minor and major axes of the ellipse were
$3.\!^{\prime\prime}12$ (60 pc) and $5.\!^{\prime\prime}22$ (100 pc)
respectively.  The X-ray emission from the two faint sources closest
to the nucleus was excluded using two $1.\!^{\prime\prime}48 \times
1.\!^{\prime\prime}48$ rectangular regions centred on their peaks.
The net (background subtracted) count rate from the nucleus was $0.33$
cts s$^{-1}$ for the short frame time observation, which does not
suffer from pile-up.  The data were initially fitted by a power-law
continuum (photon spectral index $\Gamma$) absorbed by a column
density ($N_{\rm H}$), where the atomic cross-sections and abundances
for the latter were taken from \citet{mm83} and \citet{ag89},
respectively.  We also included a Gaussian-shaped line to represent
the Fe K$\alpha$ emission.  This model (model 1) provides an
unacceptable fit with $\chi^{2} = 214.4$ for 127 degrees of freedom
(d.o.f.), the best-fitting parameters being $\Gamma =
0.02^{+0.13}_{-0.14}$ and $N_{\rm H} = 1.5\pm0.6 \times 10^{21}$
cm$^{-2}$ (the errors here, and elsewhere in this paper, are 90\%
confidence for one interesting parameter, $\Delta\chi^{2} = 2.7$).  We
note that the column density in this model description is below the
Galactic value of $N_{\rm H} (\rm Gal) = 3 \times 10^{21}$ cm$^{-2}$
\citep{fre77}, indicating the model is unphysical.  Fixing the column
density at the Galactic value worsens the fit by $\Delta\chi^{2} =
11.6$ (1 d.o.f.)  and gives a larger photon index of $\Gamma \simeq
0.22$.  Thus, we conclude that there is evidence in the ACIS data for
soft X-ray emission in excess of the absorbed power-law (cf. Matt et
al. 1996; Sako et al. 2000).  The energy and equivalent width of the
Fe K$\alpha$ emission line are $6.38$ keV and $\sim 3$ keV,
respectively.  A hard ($\Gamma \sim 0$) power-law continuum and strong
Fe K$\alpha$ emission were found in earlier \emph{ASCA}
\citep{mat96,sak00} and \emph{BeppoSAX} \citep{gua99} observations,
and in the \emph{Chandra} HETGS observation \citep{sam01b}.  These
properties are expected if we see the nuclear spectrum Compton
reflected from the far side of a putative dense torus, and do not
directly view the nucleus itself.

In addition to the Fe K$\alpha$ line at 6.38 keV, prominent emission
lines are visible in the spectrum at $\sim 0.9$, 1.2--1.3, 1.7--1.9,
2.3--2.5, and $\sim 7.0$ keV (Figure~\ref{fig5}), corresponding to the
neutral or ionized states of Ne, Mg, Si, S, and Fe K$\beta$. A large
number of emission lines have been identified in the HETGS observation
of Circinus \citep{sam01b}.  These data have much higher spectral
resolution than ours, and we decided to include the lines detected by
\citet{sam01b} in our spectral model.  Thus, the next step was to fit
the data to a model (model 2) consisting of the Compton reflected
continuum expected from a power-law spectrum (of photon index $\Gamma
= 1.7$, a typical value for an unobscured Seyfert nucleus; cf. Matt et
al. 1996; Guainazzi et al. 1999) incident on neutral, solar abundance
material (as implemented in the {\sc pexrav\/} model; see Magdziarz \&
Zdziarski 1995), a broad Gaussian-shaped line representing Fe
K$\alpha$ emission, and narrow ($\sigma = 10$~eV) Gaussian emission
lines at the energies given by \citet{sam01b}.  The inclination (to
our line of sight) of the reflecting material was fixed at the
angle-averaged value ($i = 63^{\circ}$), since the reflection spectrum
below 10 keV is relatively independent of the inclination angle.  The
shape of the incident spectrum above 10 keV is important, even when
considering measurements below this energy, because of the effects of
Compton down-scattering; the power-law continuum was therefore taken
to be exponentially cut-off with an e-folding energy of 150~keV,
similar to that observed in other Seyfert galaxies
\citep{zdz95,gon96}.  Finally, we added a thermal bremsstrahlung
continuum to represent the soft X-ray emission (a power-law continuum
of large photon index could have been used instead).  All spectral
components were assumed to be absorbed by the Galactic column density
and a column intrinsic to the nucleus.

Model 2 (Table~\ref{tbl-2}) gives an acceptable fit to the data for an
intrinsic column of $N_{\rm H} = 4.4^{+4.7}_{-2.1} \times 10^{21}$
cm$^{-2}$ and a thermal bremstrahlung temperature of $kT_{\rm brems} =
0.52^{+0.41}_{-0.29}$ keV ($\chi^{2} = 114.7$, 99 d.o.f.).  The
temperature of the thermal component agrees well with that found
($0.69^{+0.05}_{-0.07}$ keV) by \citet{sak00} in their collisionally
ionized equilibrium model.  The best-fitting energy, intrinsic width,
and normalization of the Fe K$\alpha$ emission line are
$6.382^{+0.011}_{-0.010}$ keV, $\sigma = 58\pm13$ eV, and
$2.93^{+0.33}_{-0.39} \times 10^{-4}$ photons cm$^{-2}$ sec$^{-1}$,
respectively.  Most of the line normalizations are consistent with
those of \citet{sam01b}, the exceptions being those at 1.471, 2.006,
4.758, 6.656, and 7.030 keV.  The total efficiency curve of the
back-illuminated S3 chip contains structures at the Al K-edge
(1.55--1.56 keV) and the M-edge of iridium ($\sim 2$ keV), which arise
from X-ray absorption in the optical blocking filter and mirror
surfaces respectively \citep{cha00}.  Thus, the lack of line emission
in the ACIS spectrum at 1.471 and 2.006 keV may be due to inaccuracies
in the response matrix at energies where the total efficiency of the
detector is changing rapidly.  The line present in the HETGS spectrum
at 4.758 keV has a low signal to noise ratio, and our upper limit is
consistent with the $1\sigma$ uncertainty of \citet{sam01b}.  The two
other lines, at 6.656 and 7.030 keV, have significantly lower fluxes
in the ACIS spectrum than in the HETGS spectrum.  However, the close
proximity of the 6.656 keV line to the strong Fe K$\alpha$
fluorescence line at 6.4 keV makes an accurate determination of the
line flux difficult, and it is possible that some of the HETGS line
flux may originate from regions outside our extraction area.  Our
measured flux of the Fe K$\beta$ line at 7.030 keV could be too low if
an Fe K-edge is present in excess of that assumed in the continuum
model.

\subsection{Large Scale X-ray Emission} \label{sec-large_scale}

There is a considerable amount of diffuse X-ray emission in the
Circinus field, extending hundreds of parsecs NE, SW and especially NW
from the nucleus (Figure~\ref{fig1}).  The NE and SW extensions follow
roughly the plane of the galaxy disk (p.a. $\simeq 25^{\circ}$;
Maiolino et al. 1998), while that to the NW is approximately
perpendicular to it.  This can be seen in Figure~\ref{fig6}, which is
a superposition of the 0.5--8~keV X-ray emission (greyscale) on
contours of an optical continuum image at 7000{\AA} \citep{mar94}.
Astrometric coordinates were assigned to the optical image through a
comparison between the positions of stars in this image and an
\emph{HST} I band image \citep{wil00}.  X-ray spectra were extracted
from elliptical-shape regions centred on the SW and NW extended
regions; we have excluded from the SW spectrum compact sources and
extended X-ray emission close to the nucleus (Figure~\ref{fig7}).
Background spectra were taken from a circular region
$40^{\prime\prime}$ SW of the nucleus, where there is no large-scale
extended X-ray emission from Circinus.  Soft, faint, diffuse X-ray
emission from our Galaxy is a possible contaminant of the background
given the low Galactic latitude of Circinus ($b=-3.\!^{\circ}8$;
Freeman et al. 1977).  However, similar results were obtained using
background spectra taken from other long ACIS-S observations in which
the discrete sources of X-ray emission had been excised\footnote{See
http://hea-www.harvard.edu/$\sim$\,$\!$maxim/axaf/acisbg/ for further
details regarding the ACIS background fields.}.  These observations
are of ``blank-sky'', were taken at the same focal plane temperature
as our Circinus observations (i.e. $-120^{\circ}$C), and the
background spectra were extracted from the same regions of the chip as
the source spectra in our Circinus observations.

In performing the spectral analysis of the extended X-ray emission, we
have used both the maximum-likelihood, or C-statistic (method 1; Cash
1979) and the $\chi^{2}$ statistic (method 2).  Method 1 does not
require the data to be re-binned, and thus avoids potential loss of
information on narrow spectral features.  However, use of the
C-statistic requires the background to be modelled (with simple
power-laws and Gaussians) rather than subtracted directly, and there
is no test for goodness of fit.  The confidence intervals are
calculated in the same way as for the $\chi^{2}$ statistic, and are
usually tighter for any given parameter (e.g. Nousek \& Shue 1989).
Method 2 requires the data to be binned so that there are a minimum of
(typically) 20 counts per spectral bin, and the background is
subtracted directly from the data.  We find similar values for the
best-fitting continuum parameters using the two methods, and so refer
to only the method 2 results in the text.  However, method 1 results
give more precise measurements on the line features, and so these
results are referred to when giving best-fit line energies,
normalizations, and equivalent widths.

\subsubsection{The NW region} \label{sec-nw}

We have modelled the extended emission in terms of combinations of hot
thermal plasma and power-law models.  A single component does not
describe the spectrum of the NW extension.  Therefore, the data were
fitted to a two-component model consisting of thermal bremsstrahlung
and a power-law continuum (model B+P).  This model gave an acceptable
fit to the data with $\chi^{2} = 28.9$ for 22 d.o.f., the best-fitting
parameters being given in Table~\ref{tbl-3}.  Several line features
are evident in the spectrum (Figure~\ref{fig8}), the most prominent
being those at $1.775^{+0.025}_{-0.025}$ (consistent with Si {\sc
vii--x} K$\alpha$) and $6.37^{+0.10}_{-0.06}$ keV (consistent with Fe
{\sc ii--xx} K$\alpha$).  The line at 1.78 keV is close in energy to
the Si K$\alpha$ fluorescence line in the background of the S3 chip
(see Table~3 of Chartas et al. 2000).  Thus, we consider whether this
line might be an artefact left over from the background subtraction or
modelling.  The (background subtracted) flux of the 1.78 keV line in
the data is $(8.8^{+4.9}_{-4.5}) \times 10^{-7}$ photons cm$^{-2}$
s$^{-1}$, whereas the flux of the Si K$\alpha$ line in the background
spectrum (scaled to the area of the NW extension) is $6.3 \times
10^{-9}$ photons cm$^{-2}$ s$^{-1}$.  Therefore, it is unlikely that
the line is a residual from the background subtraction or modelling.
Replacing the thermal bremsstrahlung continuum with a {\sc mekal}
model \citep{mew95} gives a slightly better fit to the data for solar
abundances (model M1+P; Table~\ref{tbl-3}).  However, the column
density is below the Galactic value so the model is unphysical.  A
significant (at $> 96$\% confidence) improvement in the fit is
obtained for a metal abundance of $0.052^{+0.047}_{-0.023}$, and a
column density consistent with the Galactic value (model M2+P;
Table~\ref{tbl-3}).  A two {\sc mekal}s (both with solar abundances)
model (model M1+M1; Table~\ref{tbl-3}) gives a similar quality fit to
the data as the M1+P model, but the column density is again well below
the Galactic value.  A significant (at $> 97$\% confidence)
improvement in the fit is obtained when we allow the metal abundance
to vary from the solar value (model M2+M2, both {\sc mekal}s have the
same metallicity, Table~\ref{tbl-3}).  However, the temperature of the
hotter {\sc mekal} component is very high ($kT_{\rm h} = 80$ keV,
which is the hottest temperature at which the model is tabulated) and
unconstrained, making it impossible to distinguish the M2+M2 from the
M2+P model.  The temperature of the cooler {\sc mekal} ($kT \sim 0.6$
keV) agrees well with the collisional ionization equilibrium model of
\citet{sak00}.  The addition of a second {\sc mekal} component to
either the M1+P or M2+P models does not significantly (at $\simgreat
70$\% confidence) improve the fit.  The best fitting models in
Table~\ref{tbl-3} have column densities consistent with the Galactic
value ($N_{\rm H} (\rm Gal) = 3 \times 10^{21}$ cm$^{-2}$).  This
result is expected since the NW region lies on the near side of the
Circinus galaxy disk, along the rotation axis.  None of the models
adequately reproduces the strength of the Fe K$\alpha$ line, whose
observed equivalent width is $2.7^{+2.4}_{-1.9}$ keV.

In the M2+P model, the best-fit power-law photon index is $\Gamma =
0.0^{+0.5}_{-1.9}$, which is close to the value observed in the
nuclear spectrum of Circinus.  Of course, the spectral index of the
hard power-law is very uncertain (Table~\ref{tbl-3}); it is consistent
with the observed value ($\Gamma \simeq 0.2$) for the nucleus, which
we interpreted as a result of reflection from an optically thick
torus, and marginally consistent ($\Delta\chi^{2} \simeq 6$) with the
putative unobscured index ($\Gamma = 1.7$).  In evaluating the
physical nature of this hard component, we first consider whether it
might result from scattering of nuclear photons by the telescope
mirrors.  From estimates of the telescope PSF at 6.4 keV, some
$0.25$\% of the photons from a point-like nucleus would fall within
the extraction region used for the NW extension.  The ratio of the
3--10 keV flux (which excludes most of the soft thermal X-ray
emission) of the NW extension to that of the nucleus is $\simeq
0.5$\%, or twice that expected from the telescope PSF.  We have also
compared the hard flux of the NW extension with the flux in a region
at the same distance from the nucleus where there is no detected
diffuse emission from Circinus (e.g. towards the SE).  The 3--10 keV
count rates of the NW extension and a region to the SE with the same
area are $(1.7 \pm 0.3) \times 10^{-3}$ cts s$^{-1}$ and $(1.2 \pm
0.2) \times 10^{-3}$ cts s$^{-1}$ respectively.  Therefore, we
tentatively conclude that the hard power-law component in the NW
extension is real. A plausible interpretation is that this hard
power-law component is nuclear flux scattered by electrons in an
ionized gas above the plane of the galaxy.  The flux and equivalent
width of the Fe K$\alpha$ emission line in the NW extension are
$(1.2^{+1.0}_{-0.8}) \times 10^{-6}$ photons cm$^{-2}$ s$^{-1}$ and
$2.7^{+2.4}_{-1.9}$ keV, respectively.  The true line flux will be
smaller than this value, because the telescope PSF contributes $7.5
\times 10^{-7}$ photons cm$^{-2}$ s$^{-1}$ from Fe K$\alpha$ emission
line in the nuclear spectrum.  However, Fe K$\alpha$ emission lines of
large equivalent width can readily occur if a small fraction of the
nuclear emission is scattered by a warm ``mirror'' \citep{kk87,ban90}.

\subsubsection{The SW region}  \label{sec-sw}

A similar spectral fitting approach was adopted for the spectrum of
the SW extension as that used for the spectrum of the NW region.
Several lines are evident in the spectrum (Figure~\ref{fig9}), the
most prominent being the Fe K$\alpha$ and K$\beta$ emission lines at
$6.410^{+0.035}_{-0.027}$ and $7.030^{+0.039}_{-0.035}$ keV
respectively.  A two-component model consisting of a thermal
bremsstrahlung plus power-law continuum (model B+P) provides a poor
fit to the data with $\chi^{2} = 42.5$ for 24 d.o.f. (the probability
that the model describes the data and that $\chi^{2}$ exceeds the
observed value by chance is $1$\%); the best-fitting parameters are
given in Table~\ref{tbl-4}.  Replacing the thermal bremsstrahlung
continuum with a solar-metallicity {\sc mekal} model gives a much
better fit to the data (model M1+P, Table~\ref{tbl-4}).  A significant
(at $> 99.5$\% confidence) improvement in the fit is obtained for a
metal abundance of $0.088^{+0.052}_{-0.031}$ times solar (model M2+P,
Table~\ref{tbl-4}).  A two {\sc mekal}s model (model M1+M1, each with
solar metallicity) gives a worse fit to the data than does the M1+P
model.  Although a significant (at $> 99.7$\% confidence) improvement
in the fit is obtained when the metal abundance is allowed to vary
(model M2+M2, both {\sc mekal}s are assumed to have the same
metallicity, Table~\ref{tbl-4}), the fit is worse than that for the
M2+P model.  Thus, we prefer the {\sc mekal} plus power-law
description of the SW extension.  It is notable that all values of the
absorbing column density in Table~\ref{tbl-4} exceed the Galactic
column, indicating that significant absorption occurs within the disk
of the Circinus galaxy.  None of the models reproduce the strong Fe
K$\alpha$ and Fe K$\beta$ lines.

Given the close proximity of the SW region to the nucleus,
approximately $1$\% of the 3--10 keV flux of a point-like nucleus
would fall within the extraction region used for the SW extension.
The ratio of the (background subtracted) 3--10 keV count rates of the
SW extension to that of the nucleus is $\simeq 1$\%, in agreement with
that expected from the telescope PSF.  Therefore, we conclude that the
hard power-law component in the SW extension is not real but
represents nuclear emission scattered by the telescope mirror.
Finally, we note that the Fe K$\alpha$ emission line flux in the SW
extension is $(2.1^{+1.6}_{-0.9}) \times 10^{-6}$ photons cm$^{-2}$
s$^{-1}$, which is also consistent with the value expected from
scattering of the strong nuclear Fe K$\alpha$ emission by the
telescope mirror.

\subsection{Compact X-ray Sources} \label{sec-compact}

Spectra were extracted using circular regions for each of the ten
non-nuclear sources listed in Table~\ref{tbl-1}.  The radius of the
extraction region was chosen to include all of the source flux above
the background level, and ranged from $1.\!^{\prime\prime}2$ to
$3.\!^{\prime\prime}0$.  Background spectra were accumulated from
annuli surrounding the source or, in the case of sources close to the
nucleus, from circular regions the same distance from the nucleus as
the source.  Prior to spectral fitting, the source spectra were
re-binned so that there were at least 20 cts bin$^{-1}$ (sources B
through K) or 10 cts bin$^{-1}$ (source A).  An absorbed power-law
continuum provides an acceptable fit to each spectrum, and the
best-fitting values of column density, $N_{\rm H}$, and power-law
photon index, $\Gamma$, are given in Table~\ref{tbl-5}.  The observed
2--10 keV fluxes range from $2.0 \times 10^{-14}$ to $5.1 \times
10^{-12}$ erg cm$^{-2}$ s$^{-1}$, corresponding to an unabsorbed X-ray
luminosity between $4.1 \times 10^{37}$ and $10^{40}$ erg s$^{-1}$,
assuming these sources are within the Circinus galaxy.  The fact that
all sources (with the possible exceptions of A, B and H) have observed
column densities in excess of the Galactic column is consistent with
them being in the Circinus galaxy.  The luminosity of the most
luminous of these sources corresponds to the Eddington luminosity of
an $80 \, M_{\odot}$ black-hole.

Significant (at $> 90$\% confidence) improvements in the fits to two
sources are obtained when we include either one or two narrow ($\sigma
= 10$ eV) Gaussian emission lines. The best-fit line energies and
observed fluxes are given in Table~\ref{tbl-6}.  The equivalent widths
are calculated with respect to the observed (absorbed) continuum.  The
presence of strong emission lines in three sources (F, G, and J) was
claimed by \citet{sam01a}, and it is worth comparing our results with
theirs (see also Bauer et al. 2001).  Source G (source 2 from Sambruna
et al. 2001a) is relatively faint, and it is therefore possible that
the lines present in our data at 1.0 and 2.8 keV correspond to the
lines at 1.2 and 2.3 keV found by \citet{sam01a}.  We note that there
is no evidence for Fe K$\alpha$ line emission in this source, which
indicates that there is little or no contamination from the nucleus.
Sources F and J are piled-up in our 3.2s frame time data, and so we
have only used the 0.4s frame time data, which has a total good time
interval of 4913s.  Our spectra (Figure~\ref{fig10}) should be of worse
quality than those obtained from the zero order image in the HETGS
observation above $\simgreat 2$ keV.  Thus, it is not surprising that
we did not detect any of the lines identified by \citet{sam01a} in
sources F and J, although we have a marginal detection of a line at
2.7 keV in source F (source 6 of Sambruna et al. 2001a).  The upper
limits from our data for the equivalent widths of the emission lines
identified by \citet{sam01a} in sources F and J are in the range $\sim
50$ to $\sim 150$ eV (and an upper limit of 1.3 keV for a line at 6.9
keV in source F is obtained from a spectral analysis of the unbinned
data using the C-statistic).

\section{Comparison with other Wavelengths} \label{sec-optical}

The [O{\sc iii}] $\lambda 5007$ and H$\alpha$ emission line images
from \citet{vb97} are compared with the X-ray image in the 0.5--8 keV
band in Figure~\ref{fig11}.  During the optical observations, the
seeing averaged FWHM $\simeq 1^{\prime\prime}$, which is similar to
the spatial resolution of the X-ray image.  As mentioned by
\citet{vb97}, the H$\alpha$ line flux was integrated over a specific
velocity range to emphasize the outflow structure and to de-emphasize
emission from the circumnuclear starburst.  Regardless, the H$\alpha$
image shows the starburst ring plus some outflowing gas, while the
[O{\sc iii}] image is dominated by the high excitation, outflowing
gas.  The large-scale diffuse X-ray emission toward the NW of the
nucleus is strongly correlated with the high excitation gas evident in
the [O{\sc iii}] image.  Thus, the X-ray emission may be associated
with a hot radiatively driven wind, shocks driven by mass outflow, or
highly excited photoionized gas (see Section~\ref{sec-discussion}).
However, the large-scale diffuse X-ray emission toward the SW is only
loosely correlated with the H$\alpha$ line emission and not at all
with the [O{\sc iii}].  This emission extends some $15^{\prime\prime}$
from the nucleus, a similar scale to the ring of H{\sc ii} regions, so
the X-ray emission here may be associated with the starburst.

\section{Discussion} \label{sec-discussion}

\subsection{Nuclear Emission}

The \emph{Chandra} ACIS spectrum of the Circinus nucleus is dominated
by a hard ($\Gamma \sim 0$) power-law continuum and strong Fe
K$\alpha$ line emission, characteristic of a Compton reflection
continuum.  A likely scenario is that the nuclear continuum, evident
in the \emph{BeppoSAX} data above $\simeq 10$ keV, is Compton
reflected from the far side of the putative dense torus.  If the
ionization cone is collimated by the torus, then the observed
half-opening angle of the cone, $\theta \simeq 45^{\circ}$
\citep{wil00}, corresponds to a torus solid angle $\Omega/4\pi = \cos
\theta = 0.7$.  This is larger than the solid angle measured from the
\emph{BeppoSAX} data ($\Omega/4\pi \simeq 0.1$; Matt et al. 1999),
which suggest that only a small fraction of the inner face of the
torus is visible directly.  For a torus solid angle of $\Omega/4\pi =
0.1$, the unobscured nuclear flux must be of order $1.2 \times
10^{42}$ erg s$^{-1}$ in the 2--10 keV band, which is within the range
of luminosities inferred from the \emph{BeppoSAX} data ($3.4 \times
10^{41}$--$1.7 \times 10^{42}$ erg s$^{-1}$; Matt et al. 1999).  Thus,
it is likely that the Compton reflection and Fe K$\alpha$ fluorescence
line observed from the nucleus originate in an almost edge-on torus.
Such an orientation would be consistent with the strong water vapor
masers in the nucleus of the Circinus galaxy (e.g. Gardner \& Whiteoak
1982).

\subsection{Large Scale X-ray Emission}

A small fraction of the hard (3--10 keV) X-ray emission in the
Circinus field extends hundreds of parsecs NW from the nucleus, and is
cospatial with the regions of high excitation gas evident in the
[O{\sc iii}] image (Figure~\ref{fig11}).  The 2--10 keV observed flux
and luminosity of this gas are $7.4 \times 10^{-14}$ erg cm$^{-2}$
s$^{-1}$ and $1.4 \times 10^{38}$ erg s$^{-1}$, respectively.  The
X-ray spectrum of the NW extension above 3 keV is similar to that
observed in the nucleus, and is marginally consistent with the
putative unobscured nuclear spectrum.  It is worthwhile to consider
whether this component could be electron scattered nuclear flux.  The
scattered X-ray luminosity, $L_{\rm scat}$, is given as
\begin{displaymath}
L_{\rm scat} = L_{\rm int} \frac{\Omega}{4\pi} \tau_{\rm scat},
\end{displaymath}
where $L_{\rm int}$ is the intrinsic nuclear luminosity, $\Omega$ is
the solid angle subtended by the scatterer as viewed from the nucleus,
and $\tau_{\rm scat}$ is the optical depth through the scattering gas.
The unabsorbed 2--10 keV nuclear luminosity is in the range $1.2
\times 10^{41}$ to $1.7 \times 10^{42}$ erg s$^{-1}$ \citep{mat99},
and the 2--10 keV luminosity of the scattered emission, after
correcting for absorption and the contribution from nuclear light
scattered by the telescope mirrors (Section~\ref{sec-nw}), is $\simeq
7 \times 10^{37}$ erg s$^{-1}$.  Assuming a half-opening angle of
$\theta \simeq 45^{\circ}$ for the ionization cone, the solid angle
subtended by the scatterer is $\Omega/4\pi = \frac{1}{2} (1 - \cos
\theta) = 0.15$, so the column density of electrons through the
scattering gas is in the range $4 \times 10^{20}$ to $2 \times
10^{21}$ cm$^{-2}$.  This is smaller than any intrinsic absorbing
column associated with the large-scale emission to the NW
(Table~\ref{tbl-3}) and thus does not require the scattering gas to be
very highly ionized for the scattered X-rays to escape.

The thermal component in the spectrum of the NW extension has a
temperature of $T \simeq 7 \times 10^{6}$ K and an emission measure of
$n_{\rm e}^{2} V \simeq 6 \times 10^{61}$ cm$^{-3}$, which yields
$n_{\rm e} \simeq 0.15$ cm$^{-3}$ assuming a volume $V = 2.8 \times
10^{63}$ cm$^{3}$ for the emitting region.  The pressure and cooling
time of the X-ray emitting gas are then $n_{\rm e} k T \simeq 1.5
\times 10^{-10}$ erg cm$^{-3}$ and $n_{\rm e} k T V / L_{\rm X} \simeq
7 \times 10^{7}$ years, respectively, assuming a 0.5--10 keV
luminosity, after correcting for absorption, of $L_{\rm X} \simeq 2.0
\times 10^{38}$ erg s$^{-1}$.  The velocity gradients across the
[O{\sc iii}] emission line images are of order $100$--$200$ km
s$^{-1}$ \citep{vb97}, indicating that the gas is too hot to have been
heated in situ by shocks with velocities in this range.  Thus, a
plausible explanation is that the X-ray emitting gas is associated
with a wind outflowing from the nucleus.  The alternative is that the
X-ray emitting gas is photo-ionized by the nucleus and thus mostly
emission lines from a much cooler plasma.  Higher spectral resolution
observations are needed to distinguish these two quite different
models.

\subsection{Compact X-ray Sources}

The compact X-ray sources within $10^{\prime\prime}$ (190 pc) of the
nucleus are most likely to be X-ray binary systems or supernova
remnants within the starburst rings of the Circinus galaxy.  The
measured column densities for at least three of the closest objects
(sources C, E, and G) are above the Galactic value, presumably because
of the large amount of molecular gas associated with the starburst
disk.  The association of the other sources with the Circinus galaxy
is less certain.  Source F is close to, but not coincident with, a
patch of H$\alpha$ emission $\simeq 27^{\prime\prime}$ south of the
nucleus, possibly an H {\sc ii} region in a spiral arm of the galaxy
\citep{elm98}.  This X-ray source has a hard X-ray spectrum and
presumably is an X-ray binary; it may show a weak extension to the SW
(Figure~\ref{fig1}).  \citet{sam01a} and \citet{bau01} report periodic
variability in source J, with the flux decreasing to zero every
$\simeq 27$ ks.  Assuming source J is within the Circinus galaxy, its
hard spectrum and high luminosity would suggest a massive ($\simgreat
80 \, M_{\odot}$) black-hole in an X-ray binary as the likely origin
for the X-ray emission.  The fact that the column density to source J
is larger than the Galactic column (Table~\ref{tbl-5}) supports an
association with Circinus.  Alternatively, if source J is within our
own galaxy, it might be associated with a magnetic Catacysmic Variable
system.

\section{Conclusions}

The main results of our \emph{Chandra} X-ray observations of the
Circinus galaxy are as follows.  
i) The nucleus contains a bright, compact X-ray source plus emission
extended by $\simeq 60$ pc in the general direction of the ionization
cone.  Our observations confirm that the observed nuclear spectrum
arises through reflection of a hard X-ray source by neutral matter
with a high column density.  Our emission-line fluxes are generally
consistent with those measured by \citet{sam01b} with much higher
spectral resolution grating observations.  The Fe K$\alpha$ emission
is extended by up to 200 pc.  
ii) There is also large scale (up to 600 pc from the nucleus),
extended emission both along the plane of the galaxy disk (i.e. to the
NE and SW) and perpendicular to it (i.e. to the NW).  The spectrum of
the extended emission along the galaxy disk may be described as
emission from gas in collisional equilibrium with $kT \sim 0.6$ keV
and may be associated with the starburst ring.  There is no
significant hard component in excess of that expected by
telescope-mirror scattering of the bright nuclear source.  The
absorbing column density to gas in the galaxy disk exceeds that from
our own galaxy, indicating significant intrinsic absorption by the
disk of Circinus.  The large-scale gas extending perpendicular to the
disk is closely correlated with the high excitation optical-line
emitting gas.  The spectrum suffers no intrinsic absorption since the
gas is on the near side of the Circinus galaxy disk.  The soft
component may be a thermal gas with $kT \sim 0.6$ keV or may be cooler
and photo-ionized by the Seyfert nucleus.  Hard emission is detected
which appears to exceed that expected from scattering of nuclear light
by the telescope mirrors.  We suggest that this hard component results
from scattering of nuclear light into our line of sight by electrons
in an ionized wind.  
iii) Ten compact X-ray sources are detected in the inner $\sim 1$ kpc
of the galaxy.  Their spectra may be described as power-laws with a
significant amount of intrinsic absorption, consistent with their
location in the inner disk of Circinus, which is known to be rich in
molecular gas.  The source luminosities range from $4 \times 10^{37}$
to $1.0 \times 10^{40}$ erg s$^{-1}$.  The most luminous source (J)
exhibits a hard spectrum and is apparently an X-ray binary in Circinus
with a black-hole mass exceeding $80 \, M_{\odot}$.

\acknowledgments

We thank Sylvain Veilleux and Alessandro Marconi for providing their
optical images of Circinus in computer readable format.  This research
made use of funding through NASA grant NAG-81027.


\clearpage



\figcaption[f1.eps]{An image of the central $1\farcm2 \times 1\farcm2$
region of the Circinus field in the 0.5--8 keV band.  The image has
been smoothed with a Gaussian profile of width $\sigma = 1 \, \rm
pixel$ ($= 0.\!^{\prime\prime}5$).  The contours indicate 2, 4, 8, 16,
64, 256, and 1024 cts pixel$^{-1}$ in the smoothed image.  Letters
refer to the sources discussed in the text and listed in
Table~\ref{tbl-1}. \label{fig1}}

\figcaption[f2.eps]{An unsmoothed image of the circumnuclear region in
the 0.5--8 keV band.  Crosses and letters mark the locations of
discrete sources found by the {\sc wavdetect} algorithm. The vertical
bar indicates the conversion from grey scale to cts pixel$^{-1}$.
\label{fig2}}

\figcaption[f3.eps]{A radial profile of the circumnuclear region in
the 0.5--8 keV band (obtained from the 0.4s frame time observation)
after excluding the contribution of compact X-ray sources.  The data
and telescope PSF are shown as the solid- and dashed-line histograms,
respectively, and 1~pixel $=0.\!^{\prime\prime}5$.  \label{fig3}}

\figcaption[f4.eps]{The distribution of Fe K$\alpha$ line emission in
the Circinus galaxy. The image has been smoothed with a Gaussian
profile of width $\sigma = 1 \, \rm pixel$ ($=0.\!^{\prime\prime}5$).
Contours are drawn at $-1$, $-0.5$ (dashed), 0.5, 1, 2, 4, 8, 16, 32,
64, and 128 cts pixel$^{-1}$ in the smoothed image. The crosses mark
the positions of the sources detected in the 0.5--8 keV band with the
{\sc wavdetect} algorithm (Section~\ref{sec-overview}). The features
in the extreme upper left (near sources I and J) reflect residuals
errors in the subtraction of the continuum rather than real Fe
K$\alpha$ emission.  \label{fig4}}

\figcaption[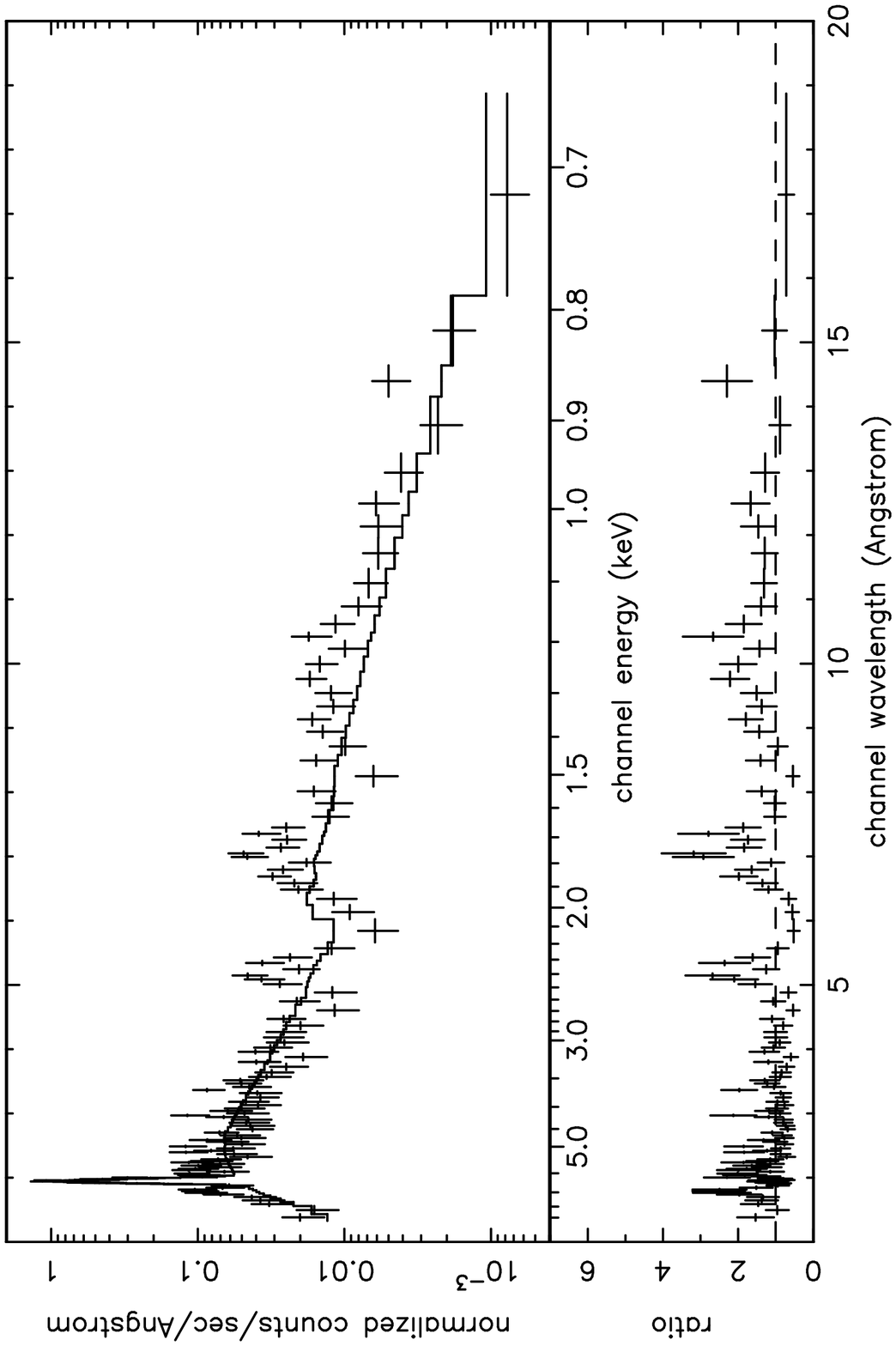]{The nuclear spectrum of Circinus.  The upper panel
shows the data and folded model comprising the best-fit absorbed
power-law continuum plus Fe K$\alpha$ emission line (model 1;
Section~\ref{sec-nuc_spectrum}). The lower panel shows the ratio of
data to the model.  Several prominent emission lines are seen in the
spectrum at energies corresponding to neutral or ionized species of
Ne, Mg, Si, S, and Fe K$\beta$ (see Table~\ref{tbl-2}).
\label{fig5}}

\figcaption[f6.eps]{An optical continuum image (contours) of the
Circinus galaxy at 7000{\AA} (from Marconi et al. 1994) superposed on
the \emph{Chandra} X-ray image (greyscale) in the 0.5--8 keV band,
smoothed with a Gaussian profile of width $\sigma = 1 \, \rm pixel$
($=0.\!^{\prime\prime}5$).  The vertical bar indicates the conversion
from grey scale to cts pixel$^{-1}$ in the X-ray image.  Contours are
drawn at 0.9, 1.1, 1.3, 1.65, 2.2, 3.3, 5.5, 11.0, and $22.0 \times
10^{-18}$ erg cm$^{-2}$ s$^{-1}$ {\AA}$^{-1}$ pixel$^{-1}$ in the
optical image. \label{fig6}}

\figcaption[f7.eps]{An image of an $\simeq 1^{\prime} \times
1^{\prime}$ region of the Circinus field in the 0.5--8 keV band.  The
image has been smoothed with a Gaussian profile of width $\sigma = 1
\, \rm pixel$ ($=0.\!^{\prime\prime}5$).  The crosses mark the
locations of the discrete sources found by the {\sc wavdetect}
algorithm.  The solid black lines show regions from which spectra of
the diffuse X-ray emission were extracted.  The vertical bar indicates
the conversion from grey scale to cts pixel$^{-1}$. \label{fig7}}

\figcaption[f8.eps]{The spectrum of the NW large scale extended
emission in Circinus.  The data are shown together with the best-fit
absorbed thermal bremsstrahlung plus power-law continuum model (model
B+P; see Table~\ref{tbl-3} and Section~\ref{sec-nw}).  The vertical
lines mark the best-fit energies of the emission lines present in the
spectrum. \label{fig8}}

\figcaption[f9.eps]{The spectrum of the SW large scale extended
emission in Circinus.  The data are shown together with the best-fit
absorbed thermal bremsstrahlung plus power-law continuum model (model
B+P; see Table~\ref{tbl-4} and Section~\ref{sec-sw}).  The emission
above 3 keV, including the observed Fe K$\alpha$ line, is believed not
to be real, but rather represents nuclear emission scattered into the
region by the telescope mirrors (Section~\ref{sec-sw}). \label{fig9}}

\figcaption[f10.eps]{Short (0.4s; triangles) and long (3.2s; crosses)
frame time spectra of two of the ten compact X-ray sources found in
the central $\sim 1^{\prime} \times 1^{\prime}$ region of the Circinus
field.  The upper panel shows the data and folded model comprising the
best-fit absorbed power-law continuum (see Table~\ref{tbl-5} and
Section~\ref{sec-compact}).  The lower panel shows the ratio of data
to the model.  The hard ($\simgreat 7$ keV) X-ray emission in excess
of the absorbed power-law is a consquence of pileup in the 3.2s frame
time data.  (a) Source F.  (b) Source J. \label{fig10}}

\figcaption[f11.eps]{Optical emission line images (from Veilleux \&
Bland-Hawthorn 1997) of the Circinus galaxy (contours) superposed on
the \emph{Chandra} X-ray image (greyscale) in the 0.5--8 keV band,
smoothed with a Gaussian profile of width $\sigma = 1 \, \rm pixel$
($=0.\!^{\prime\prime}5$). The vertical bar represents the relation
between shade and cts pixel$^{-1}$ in the X-ray image. (a) [O{\sc
iii}] $\lambda 5007$.  Contours represent 50, 150, 300, 600, and 1200
cts pixel$^{-1}$.  (b) Blueshifted (between $-150$ and $0$ km
s$^{-1}$) H$\alpha$. Contours represent 25, 50, 100, 250, 500, 1000,
and 2500 cts pixel$^{-1}$. \label{fig11}}

\clearpage

\begin{deluxetable}{cccc}
\footnotesize
\tablecaption{Bright sources in the inner part of the Circinus field. 
\label{tbl-1}}
\tablewidth{0pt}
\tablehead{
\colhead{Source} &
\colhead{RA(2000)} & 
\colhead{Dec(2000)} &
\colhead{Count rate\tablenotemark{a}}}
\startdata
A & 14 13 05.54 & -65 20 31.6 & $1.9\pm0.3$  \\
B & 14 13 09.15 & -65 20 17.7 & $5.3\pm0.6$  \\
C & 14 13 09.60 & -65 20 22.1 & $3.2\pm0.8$  \\
D & 14 13 09.94 & -65 20 21.0 & $330\pm8$    \\
E & 14 13 10.00 & -65 20 29.8 & $6.0\pm0.6$  \\
F & 14 13 10.05 & -65 20 44.8 & $102\pm5$    \\ 
G & 14 13 10.29 & -65 20 18.0 & $25.9\pm1.2$ \\
H & 14 13 10.37 & -65 20 22.7 & $8.8\pm0.7$  \\
I & 14 13 12.20 & -65 20 07.4 & $10.5\pm0.7$ \\
J & 14 13 12.25 & -65 20 14.0 & $341\pm9$    \\
K & 14 13 12.54 & -65 20 52.8 & $6.9\pm0.5$  \\
\enddata

\tablenotetext{a}{0.1--10 keV count rate expressed in units of
$10^{-3}$ counts s$^{-1}$}


\end{deluxetable}

\clearpage

\begin{deluxetable}{llcccc} 
\footnotesize 
\setlength{\tabcolsep}{0.02in} 
\tablecolumns{6} 
\tablewidth{0pt}
\tablecaption{Emission lines from the nucleus. \label{tbl-2}} 
\tablehead{ 
\colhead{Energy\tablenotemark{a}} & \colhead{Line(s)} & 
\colhead{Flux\tablenotemark{b}} & \colhead{EW\tablenotemark{c}} & 
\colhead{Flux\tablenotemark{d}} & \colhead{Ratio\tablenotemark{e}} \\
\colhead{(keV)} & \colhead{} & 
\colhead{($10^{-5}~\rm ph~cm^{-2}~s^{-1}$)} & \colhead{(eV)} & 
\colhead{($10^{-5}~\rm ph~cm^{-2}~s^{-1}$)} & \colhead{}}
\startdata 
1.211 & Ne {\sc x} Ly$\beta$ & $1.35^{+1.3}_{-1.2}$ & $130^{+130}_{-110}$ & $0.25\pm0.15$ & $5.4\pm6.2$ \\
1.332 & Mg {\sc xi} f & $0.3^{+1.6}_{-0.3}$ & $30^{+160}_{-30}$ & $0.34\pm0.16$ & $0.85\pm4.7$ \\
1.341 & Mg {\sc xi} i & $<2.0$ & $<215$ & $0.26\pm0.13$ & $\simless 7.7$ \\
1.352 & Mg {\sc xi} r & $0.8^{+1.2}_{-0.8}$ & $80^{+130}_{-80}$ & $0.58\pm0.17$ & $1.3\pm2.1$ \\
1.471 & Mg {\sc xii} Ly$\alpha$ & $<0.43\tablenotemark{f}$ & $<49$ & $0.65\pm0.16$ & $\simless 0.66\tablenotemark{f}$ \\
1.578 & Mg {\sc xi} 1s3p--1s$^{2}$ & $0.22^{+0.66}_{-0.22}$ & $28^{+84}_{-28}$ & $0.19\pm0.13$ & $1.2\pm3.6$ \\
1.741 & Mg {\sc xii} Ly$\beta$ & $1.89^{+0.92}_{-0.69}$ & $260^{+130}_{-100}$ & $0.86\pm0.17$ & $2.2\pm1.2$ \\
 & + Si {\sc ii--vi} K$\alpha$ & & & & \\
1.841 & Si {\sc xiii} f & $<1.5$ & $<230$ & $1.05\pm0.18$ & $\simless 1.4$ \\
 & + Mg {\sc xii} Ly$\gamma$ & & & & \\
1.863 & Si {\sc xiii} r & $1.6^{+0.9}_{-1.6}$ & $265^{+140}_{-255}$ & $0.87\pm0.16$ & $1.9\pm1.8$ \\
2.006 & Si {\sc xiv} Ly$\alpha$ & $<0.37\tablenotemark{f}$ & $<63$ & $0.64\pm0.15$ & $\simless 0.58\tablenotemark{f}$ \\
2.313 & S {\sc ii--xii} K$\alpha$ & $1.56^{+0.81}_{-0.82}$ & $270\pm140$ & $0.29\pm0.32$ & $5.4\pm6.6$ \\
2.383 & Si {\sc xiv} Ly$\beta$ & $<1.0$ & $<170$ & $0.48\pm0.37$ & $\simless 2.1$ \\
 & + S {\sc xii} 1s2s$^{2}$2p$^{2}$--1s$^{2}$2s$^{2}$2p & & & & \\ 
2.412 & S {\sc xiv} 1s2s2p--1s$^{2}$2s & $<1.5$ & $<260$ & $0.28\pm0.27$  & $\simless 5.4$ \\
2.433 & S {\sc xv} f & $1.2^{+1.5}_{-1.2}$ & $200^{+260}_{-200}$ & $0.56\pm0.29$ & $2.1\pm2.9$ \\
2.458 & S {\sc xv} r & $0.9^{+1.5}_{-0.9}$ & $150^{+260}_{-150}$ & $0.85\pm0.36$ & $1.0\pm1.8$ \\
2.624 & S {\sc xvi} Ly$\alpha$ & $0.36^{+0.58}_{-0.36}$ & $60^{+100}_{-60}$ & $0.22\pm0.23$ & $1.6\pm3.1$ \\
2.882 & S {\sc xv} 1s3p-1s$^{2}$ & $0.23^{+0.57}_{-0.23}$ & $36^{+89}_{-36}$ & $0.21\pm0.23$ & $1.1\pm3.0$ \\
2.960 & Ar {\sc ii--xi} K$\alpha$ & $0.46^{+0.60}_{-0.46}$ & $69^{+90}_{-69}$ & $0.6\pm0.31$  & $0.8\pm1.1$ \\
3.268 & Ar {\sc xviii} Ly$\alpha$ & $0.10^{+0.52}_{-0.10}$ & $14^{+73}_{-14}$ & $0.40\pm0.24$ & $0.25\pm1.3$ \\
3.683 & Ca {\sc ii--xiv} K$\alpha$ & $0.76^{+0.59}_{-0.60}$ & $92^{+71}_{-73}$ & $0.38\pm0.21$ & $2.0\pm1.9$ \\
3.917 & Ar {\sc xviii} Ly$\beta$ & $<0.50$ & $<57$ & $0.41\pm0.26$  & $\simless 1.2$ \\
4.758 & & $<0.28$ & $<27$ & $0.52\pm0.29$ & $\simless 0.54$ \\
5.166 & & $0.22^{+0.71}_{-0.22}$ & $19^{+61}_{-19}$ & $1.02\pm0.40$ & $0.22\pm0.70$ \\
5.993 & & $0.49^{+0.96}_{-0.49}$ & $38^{+74}_{-38}$ & $1.45\pm0.54$ & $0.34\pm0.67$ \\
6.382\tablenotemark{g} & Fe {\sc ii-xvii} K$\alpha$ & $29.4^{+3.3}_{-3.9}$ & $2250^{+260}_{-300}$ & $30.5\pm1.7$ & $0.96\pm0.14$ \\
6.656 & Fe {\sc xxv} r & $<0.94$ & $<69$ & $2.73\pm1.10$ & $\simless 0.34$ \\
7.030 & Fe {\sc ii--xvii} K$\beta$ & $2.25\pm1.7$ & $210\pm160$ & $6.8\pm1.5$ & $0.33\pm0.26$ \\
\enddata 


\end{deluxetable}

\clearpage

\setlength{\parindent}{0pt}

$^{a}$Emission line energy measured by \citet{sam01b} \\
$^{b}$Line flux from the present data \\
$^{c}$Equivalent width with respect to the observed continuum from the
present data \\
$^{d}$Line flux from the HETGS spectrum \citep{sam01b} \\
$^{e}$Ratio of the ACIS (present data) to HETGS \citep{sam01b} line
fluxes; the larger error on the ACIS flux was used in calculating the
uncertainty on the line ratio \\
$^{f}$Unreliable measurement due to structure in the instrumental
response (see text) \\
$^{g}$Line energy measured in ACIS data \\

\clearpage

\begin{deluxetable}{ccccccccc} 
\footnotesize
\rotate
\tablecaption{Spectral fits to the NW large-scale extended emission. \label{tbl-3}} 
\tablewidth{0pt} 
\tablehead{ 
\colhead{Model\tablenotemark{a}} & 
\colhead{Method\tablenotemark{b}} &
\colhead{$kT_{\rm l}$\tablenotemark{c}} &
\colhead{$kT_{\rm h}$\tablenotemark{d}} & 
\colhead{$Z$\tablenotemark{e}} & 
\colhead{$\Gamma$\tablenotemark{f}} & 
\colhead{$N_{\rm H}$\tablenotemark{g}} & 
\colhead{$F_{\rm X}$\tablenotemark{h}} &
\colhead{$\chi^{2}$ (d.o.f.)} \\
\colhead{} &
\colhead{} &
\colhead{(keV)} &
\colhead{(keV)} &
\colhead{($Z_{\odot}$)} &
\colhead{} &
\colhead{($10^{21}$ cm$^{-2}$)} &
\colhead{($10^{-14}$ erg cm$^{-2}$ s$^{-1}$)} &
\colhead{}}
\startdata 
B+P & 1 & $0.187^{+0.090}_{-0.053}$ & ... & ... & $1.96^{+0.47}_{-0.50}$ & 
	$6.7^{+0.6}_{-1.4}$ & $4.8$ & ... \\
B+P & 2 & $0.137^{+0.036}_{-0.028}$ & ... & ... & $2.60^{+0.52}_{-0.65}$ & 
	$8.7^{+0.45}_{-0.28}$ & $3.4$ & 28.9 (22) \\
M1+P & 1 & ... & $0.641^{+0.058}_{-0.066}$ & $1.0^{\rm fixed}$ & 
	$2.10^{+0.38}_{-0.39}$ & $1.63^{+0.68}_{-0.70}$ & $4.2$ & ... \\
M1+P & 2 & ... & $0.630^{+0.063}_{-0.083}$ & $1.0^{\rm fixed}$ & 
	$2.33^{+0.55}_{-0.27}$ & $1.88^{+0.72}_{-0.72}$ & $3.3$ & 25.8 (22) \\
M2+P & 1 & ... & $0.63^{+0.08}_{-0.10}$ & $0.054^{+0.048}_{-0.023}$ & 
	$0.4^{+1.1}_{-1.2}$ & $2.34^{+0.61}_{-0.57}$ & $7.1$ & ... \\
M2+P & 2 & ... & $0.63^{+0.08}_{-0.11}$ & $0.052^{+0.047}_{-0.023}$ & 
	$0.0^{+0.5}_{-1.9}$ & $2.36^{+0.60}_{-0.56}$ & $7.4$ & 20.6 (21) \\
M1+M1 & 1 & $0.664^{+0.097}_{-0.055}$ & $5.3^{+4.1}_{-1.7}$ & 
	$1.0^{\rm fixed}$ & ... & $0.74^{+0.45}_{-0.39}$ & $5.0$ & ... \\
M1+M1 & 2 & $0.655^{+0.053}_{-0.056}$ & $4.1^{+2.3}_{-1.2}$ & 
	$1.0^{\rm fixed}$ & ... & $0.72^{+0.41}_{-0.38}$ & $3.8$ & 28.6 (22) \\
M2+M2 & 1 & $0.61^{+0.09}_{-0.11}$ & $80_{-71}$ & $0.065^{+0.067}_{-0.028}$ & 
	... & $2.21^{+0.68}_{-0.76}$ & $5.0$ & ... \\
M2+M2 & 2 & $0.62^{+0.07}_{-0.14}$ & $80_{-75}$ & $0.068^{+0.049}_{-0.032}$ & 
	... & $2.12^{+0.84}_{-0.63}$ & $4.4$ & 22.4 (21) \\
\enddata 


\end{deluxetable} 

\clearpage

\setlength{\parindent}{0pt}

$^{a}$Model: B = bremsstrahlung; P = power-law; M1 = {\sc mekal} with
solar abundances (model valid for gas temperatures in the range 8 eV
to 80 keV); M2 = {\sc mekal} with variable abundances (model valid for
gas temperatures in the range 8 eV to 80 keV) \\
$^{b}$1: spectral analysis of unbinned data using the C-statistic
\citep{cas79}; 2: spectral analysis of binned data using the
$\chi^{2}$ statistic \\
$^{c}$Low temperature thermal component \\
$^{d}$High temperature thermal component \\
$^{e}$Metal abundance\\
$^{f}$Photon spectral index \\
$^{g}$Measured column density \\
$^{h}$Observed 2--10 keV flux \\

\clearpage

\begin{deluxetable}{ccccccccc} 
\footnotesize
\rotate
\tablecaption{Spectral fits to the SW large-scale extended emission. \label{tbl-4}} 
\tablewidth{0pt} 
\tablehead{ 
\colhead{Model\tablenotemark{a}} & 
\colhead{Method\tablenotemark{b}} &
\colhead{$kT_{\rm l}$\tablenotemark{c}} &
\colhead{$kT_{\rm h}$\tablenotemark{d}} & 
\colhead{$Z$\tablenotemark{e}} & 
\colhead{$\Gamma$\tablenotemark{f}} & 
\colhead{$N_{\rm H}$\tablenotemark{g}} & 
\colhead{$F_{\rm X}$\tablenotemark{h}} &
\colhead{$\chi^{2}$ (d.o.f.)} \\
\colhead{} &
\colhead{} &
\colhead{(keV)} &
\colhead{(keV)} &
\colhead{($Z_{\odot}$)} &
\colhead{} &
\colhead{($10^{21}$ cm$^{-2}$)} &
\colhead{($10^{-14}$ erg cm$^{-2}$ s$^{-1}$)} &
\colhead{}}
\startdata 
B+P & 1 & $0.126^{+0.060}_{-0.029}$ & ... & ... & $3.03^{+0.25}_{-0.18}$
	 & $12.88^{+0.76}_{-0.48}$ & $4.6$ & ... \\
B+P & 2 & $0.099^{+0.030}_{-0.022}$ & ... & ... & $4.04^{+0.24}_{-0.20}$
	 & $15.15^{+0.57}_{-0.52}$ & $3.0$ & 42.5 (24) \\
M1+P & 1 & ... & $0.628^{+0.053}_{-0.051}$ & $1.0^{\rm fixed}$ & 
	$2.30^{+0.46}_{-0.50}$ & $5.41^{+0.97}_{-0.89}$ & $5.1$ & ... \\
M1+P & 2 & ... & $0.653^{+0.091}_{-0.063}$ & $1.0^{\rm fixed}$ & 
	$2.95^{+0.49}_{-0.44}$ & $5.56^{+0.85}_{-0.77}$ & $3.3$ & 30.5 (24) \\
M2+P & 1 & ... & $0.643^{+0.059}_{-0.057}$ & $0.099^{+0.052}_{-0.038}$ & 
	$-1.9^{+1.2}_{-1.5}$ & $5.74^{+0.89}_{-0.78}$ & $18$ & ... \\
M2+P & 2 & ... & $0.68^{+0.14}_{-0.06}$ & $0.088^{+0.052}_{-0.031}$ & 
	$-5.4^{+0.3}_{-4.0}$ & $5.42^{+0.91}_{-0.11}$ & $76$ & 21.4 (23) \\
M1+M1 & 1 & $0.623^{+0.047}_{-0.043}$ & $11^{+32}_{-5}$ & 
	$1.0^{\rm fixed}$ & ... & $6.2^{+1.1}_{-1.2}$ & $6.8$ & ... \\
M1+M1 & 2 & $0.66^{+0.10}_{-0.05}$ & $2.9^{+2.9}_{-0.7}$ & $1.0^{\rm fixed}$ &
	 ... & $4.9^{+1.9}_{-0.6}$ & $4.1$ & 42.6 (24) \\
M2+M2 & 1 & $0.618^{+0.060}_{-0.066}$ & $80_{-65}$ & $0.14^{+0.10}_{-0.06}$
	 & ... & $5.60^{+0.91}_{-0.84}$ & $6.0$ & ... \\
M2+M2 & 2 & $0.65^{+0.10}_{-0.06}$ & $80_{-78}$ & $0.109^{+0.083}_{-0.042}$ 
	& ... & $5.34^{+0.87}_{-0.85}$ & $4.6$ & 28.1 (23) \\
\enddata 


\end{deluxetable} 

\clearpage

\setlength{\parindent}{0pt}

$^{a}$Model: B = bremsstrahlung; P = power-law; M1 = {\sc mekal} with
solar abundances (model valid for gas temperatures in the range 8 eV
to 80 keV); M2 = {\sc mekal} with variable abundances (model valid for
gas temperatures in the range 8 eV to 80 keV) \\
$^{b}$1: spectral analysis of unbinned data using the C-statistic
\citep{cas79}; 2: spectral analysis of binned data using the
$\chi^{2}$ statistic \\
$^{c}$Low temperature thermal component \\
$^{d}$High temperature thermal component \\
$^{e}$Metal abundance \\
$^{f}$Photon spectral index \\
$^{g}$Measured column density \\
$^{h}$Observed 2--10 keV flux \\

\clearpage

\begin{deluxetable}{ccccccc} 
\footnotesize
\rotate
\tablecaption{Spectral fits to the bright sources in the field. \label{tbl-5}} 
\tablewidth{0pt} 
\tablehead{ 
\colhead{Source} &
\colhead{$A$\tablenotemark{a}} & 
\colhead{$\Gamma$\tablenotemark{b}} & 
\colhead{$N_{\rm H}$\tablenotemark{c}} & 
\colhead{$F_{\rm X}$\tablenotemark{d}} & 
\colhead{$L_{\rm X}$\tablenotemark{e}} &
\colhead{$\chi^{2}$ (d.o.f.)} \\
\colhead{} &
\colhead{($10^{-6}$ ph cm$^{-2}$ sec$^{-1}$)} &
\colhead{} &
\colhead{($10^{21}$ cm$^{-2}$)} &
\colhead{($10^{-14}$ erg cm$^{-2}$ s$^{-1}$)} &
\colhead{($10^{37}$ erg s$^{-1}$)} &
\colhead{}}
\startdata 
A & $6.1$ & $1.8^{+1.3}_{-0.9}$ & $8^{+12}_{-5}$ & $2.0$ & $4.1$ & 0.89 (3) \\
B & $8.8$ & $0.84^{+0.96}_{-0.76}$ & $13^{+11}_{-10}$ & $14$ & $28$ & 5.6 (6) \\ 
C & $44$ & $2.6^{+1.9}_{-1.3}$ & $20^{+20}_{-12}$ & $4.0$ & $9.7$ & 14.6 (12) \\ 
E & $89$ & $2.45^{+0.94}_{-0.66}$ & $31^{+16}_{-12}$ & $8.8$ & $23$ & 2.2 (9) \\ 
F & $280$ & $1.62^{+0.32}_{-0.26}$ & $9.7^{+3.0}_{-2.0}$ & $120$ & $230$ & 28.2 (20) \\
G & $310$ & $2.85^{+0.43}_{-0.33}$ & $16.6^{+4.1}_{-3.0}$ & $19$ & $46$ & 41.5 (39) \\
H & $7.8$ & $0.45^{+0.39}_{-0.31}$ & $7.5^{+5.4}_{-3.6}$ & $26$ & $51$ & 8.5 (12) \\ 
I & $23$ & $1.19^{+0.40}_{-0.27}$ & $11.8^{+5.8}_{-3.8}$ & $20$ & $41$ & 10.6 (13) \\
J & $820$ & $1.39^{+0.15}_{-0.14}$ & $9.8^{+1.4}_{-1.1}$ & $510$ & $1000$ & 66.5 (70) \\
K & $31$ & $1.84^{+0.70}_{-0.34}$ & $14.1^{+8.2}_{-4.6}$ & $9.0$ & $20$ & 7.2 (7) \\
\enddata 


\tablenotetext{a}{Photon flux of the unabsorbed power-law continuum at 1 keV}
\tablenotetext{b}{Photon spectral index}
\tablenotetext{c}{Measured column density}
\tablenotetext{d}{Observed 2--10 keV flux}
\tablenotetext{e}{Unobscured 2--10 keV luminosity, assuming the source
is in the Circinus galaxy (assumed distance $= 4~\rm Mpc$)}

\end{deluxetable} 

\clearpage

\begin{deluxetable}{cccccc} 
\footnotesize
\tablecaption{Emission lines from the bright sources in the field. \label{tbl-6}} 
\tablewidth{0pt} 
\tablehead{ 
\colhead{Source} &
\colhead{Energy\tablenotemark{a}} & 
\colhead{Line(s)\tablenotemark{b}} & 
\colhead{Flux\tablenotemark{c}} & 
\colhead{EW\tablenotemark{d}} &
\colhead{$\Delta\chi^{2}$ (d.o.f.)} \\
\colhead{} &
\colhead{(keV)} &
\colhead{} &
\colhead{($10^{-6}$ ph cm$^{-2}$ s$^{-1}$)} &
\colhead{(eV)} &
\colhead{}}
\startdata 
G & $1.006^{+0.032}_{-0.028}$ & Ne {\sc x} Ly$\alpha$ & 
	$1.18^{+0.76}_{-0.73}$ & $290^{+190}_{-180}$ & 6.5 (2) \\ 
G & $2.773^{+0.031}_{-0.039}$ & S {\sc xv} 1s3p--1s$^{2}$ & 
	$2.4^{+0.9}_{-1.3}$ & $200^{+80}_{-110}$ & 11.0 (2) \\ 
I & $4.02^{+0.09}_{-0.65}$ & Ca {\sc ii--xx} & $1.06^{+0.89}_{-0.93}$ 
	& $280^{+240}_{-250}$ & 3.6 (2) \\
\enddata 


\tablenotetext{a}{Emission line energy measured in the ACIS data}
\tablenotetext{b}{Tentative identification}
\tablenotetext{c}{Line flux}
\tablenotetext{d}{Equivalent width with respect to the observed continuum}

\end{deluxetable} 

\end{document}